\documentstyle[12pt]{article}

\begin{document}

\baselineskip=18.6pt plus 0.2pt minus 0.1pt


\begin{titlepage}
\title{
\hfill\parbox{4cm} {\normalsize UFR-HEP/01-11}\\ \vspace{1cm}
     {\bf      On Non Commutative Calabi-Yau Hypersurfaces  }
}
\author{
A. Belhaj\thanks{{\tt ufrhep@fsr.ac.ma}}  and E.H.
Saidi\thanks{{\tt H-saidi@fsr.ac.ma}} {}
\\[7pt]
{\it UFR-High Energy Physics Laboratory, Faculty of Sciences,
Rabat, Morocco.} }

\maketitle \thispagestyle{empty}
\begin{abstract}
Using the algebraic geometry method of Berenstein et \textit{al}
(hep-th/0005087), we reconsider the derivation of the non
commutative quintic algebra  ${\mathcal{A}}_{nc}(5)$ and derive
new representations by choosing different sets of Calabi-Yau
charges $\left\{ C_{i}^{a}\right\} $. Next we extend these results
to higher $d$ complex dimension non commutative Calabi-Yau
hypersurface algebras ${\mathcal{A}}_{nc}(d+2)$. We derive and
solve the set of constraint eqs carrying the non commutative
structure in terms of Calabi-Yau
charges and discrete torsion. Finally we construct the representations
of ${\mathcal{A}}%
_{nc}(d+2) $ preserving manifestly the Calabi-Yau condition $%
\sum_{i}C_{i}^{a}=0$ and give comments on the non commutative
subalgebras.
\end{abstract}

 \newpage

\end{titlepage}
\newpage
\def\be{\begin{equation}}
\def\ee{\end{equation}}
\def\bea{\begin{eqnarray}}
\def\eea{\end{eqnarray}}
\def\nn{\nonumber}
\def\l{\lambda}
\def\t{\times}
\def\[{\bigl[}
\def\]{\bigr]}
\def\({\bigl(}
\def\){\bigr)}
\def\p{\partial}
\def\o{\over}
\def\ta{\tau}
\def\cm{\cal M}
\def\R{\bf R}
\def\b{\beta}
\def\a{\alpha}

\section{Introduction}

Since the original work of Connes et \textit{al} on Matrix model
compactification on non commutative (NC) torii \cite{1}, an
increasing interest has been devoted to the study of non
commutative spaces in connection with solitons in NC quantum field
\cite{2,3}, and string field theories \cite{4}. These NC solitons,
which have been subject to an intensive interest during the last
few years, are involved in the study of $D(p-4)/Dp$ brane systems
($p>3$) of superstrings; in particular in the ADH construction of
the $D0/D4$ system \cite{5}, in the determination of the vacuum
field solutions of the Higgs branch of supersymmetric gauge
theories with eight supercharges \cite{6,7} and in tachyon
condensation using the GMS approach \cite{8}. However most of the
NC spaces
considered in these studies involve mainly NC $R_{\theta }^{d}$ , NC $%
T_{\theta }^{d}$ torii \cite{9,10}, some cases of $Z_{n}$ type
orbifolds of NC torii \cite{11,12} and some generalizations to non
commutative higher dimensional cycles such as the non commutative
Hirzebruch complex surface $F_{0}$ used in \cite{13}. \par Quite
recently efforts have been devoted to go beyond these particular
manifolds. A special interest has been given to build non
commutative Calabi-Yau manifolds containing the commutative ones
as subalgebras and a development has been obtained for the case of
orbifolds of Calabi-Yau hypersurfaces with discrete torsion. The
key point of this construction, using a NC algebraic geometric
method \cite{14}, is based on solving non commutativity in terms
of discrete torsion of the orbifolds \cite{15}. More precisely,\
it has been shown that the $\frac{T^{2}\times T^{2}\times
T^{2}}{\mathbf{{Z_{2}}\times {Z_{2}}}}$\ orbifold of the three
elliptic curves with torsion, embedded in the $C^{6}$ complex
space, define a noncommutative Calabi-Yau threefolds where non
commutativity, carried by the discrete torsion, has a remarkable
interpretation in terms of closed string states. Moreover on the
fixed planes of this non commutative threefolds, branes
fractionate and local deformations are no more trivial. This idea
was successfully applied for orbifolds of the quintic Calabi-Yau
threefolds. In this regards it was shown
that the local non commutative quintic, to which we refer herebelow to as $%
{\mathcal{A}}_{nc}(5),$ is a subalgebra of $Mat(5,C)$, the algebra
of $5\times 5$\ complex matrices. The ${\mathcal{A} }_{nc}(5)$
reads as:
\begin{eqnarray}
Z_{1}Z_{2} &=\alpha Z_{2}Z_{1},\qquad Z_{1}Z_{3}&=\alpha
^{-1}\beta Z_{3}Z_{1},   \nn\\ Z_{1}Z_{4} &=\beta
^{-1}Z_{4}Z_{1},\qquad Z_{2}Z_{3}&=\alpha \gamma Z_{3}Z_{2}, \nn\\
Z_{2}Z_{4} &=\gamma ^{-1}Z_{4}Z_{2},\qquad Z_{3}Z_{4}&=\beta
\gamma Z_{4}Z_{3},
 \\ Z_{i}Z_{5} =&Z_{5}Z_{i},\qquad i =1,2,3,4;\nn
\end{eqnarray}
where $\alpha ,\beta $\ and\ $\gamma $\ are fifth roots of the
unity generating the $\mathbf{Z}_{5}^{3}$\ discrete group and
where the $Z_{i}$\ 's are the generators of
${\mathcal{A}}_{nc}(5)$. One of the main features of this non
commutative algebra is that its centre $\mathcal{Z}$ coincides
exactly with the homogeneous polynomial of the quintic Calabi-Yau
threefolds. \\In this study we reconsider the derivation of this
algebra by using the non commutative algebraic geometry approach
of \cite{15} and reformulate the non commutativity structure in
terms of set of constrains compatible with the Calabi-Yau
condition and discrete torsion. Then we construct new
representations of ${\mathcal{A}}_{nc}(5)$. One  of the
interesting results we have got here is that to  each  set of
Calabi-Yau charges $\left\{
C_{i}^{a}\right\} ,$ it is  associated a representation ${\mathcal{R}}_{\{{C_{i}^{a}}\}}$%
 of the non commutative algebra ${\mathcal{A}}_{nc}(5)$. We also
give the general solution for the
${\mathcal{R}}_{\{{C_{i}^{a}}\}}$\ matrix representations of
${\mathcal{A}}_{nc}(5))$ preserving manifestly the Calabi-Yau
condition and generalizing that derived in \cite{15}. Next we
extend these results to higher dimensions by deriving the $d$
complex non commutative orbifolds
of Calabi-Yau hypersurfaces with discrete torsion, ${\mathcal{A}}_{nc}(d+2)$%
\thinspace\ as well as their representations. We discuss bulk
representations and fractional branes at orbifolds points. \par
The organization of this paper is as follows: In section 2, we
reconsider the study of the non commutative quintic and give new
representations. Then we give the analogue of the NC quintic in
weighted projective space $WP^{4}$. In section 3, we explore the general form of the non commutative algebra $%
{\mathcal{A}}_{nc}(d+2)$ associated with complex Calabi-Yau
hypersurfaces with discrete torsion embedded in
$\mathbf{WP}^{d+1}$ while in section 4 we give the explicit form
of its representations for both regular and singular points. Our
solutions depends manifestly of on the data of the Calabi-Yau. We
end this paper by giving our conclusion.

\section{Local Non Commutative Quintic}

To build non commutative quintic according to \cite{14,15}, one
starts from the complex homogeneous hypersurface
\begin{equation}
P_{5}(z_{1},...,z_{5})=z_{1}^{5}+z_{2}^{5}+z_{3}^{5}+z_{4}^{5}+z_{5}^{5}+a_{0}\prod_{i=1}^{5}z_{i}=0,
\end{equation}
where $(z_{1},z_{2},z_{3},z_{4},z_{5})$ are the homogeneous coordinates of $%
\mathbf{P}^{4}$ and $a_{0}$ is a complex parameter. This
polynomial has a set of discrete isometries acting on the
homogeneous coordinates $z_{i}$ as:
\begin{equation}
z_{i}\rightarrow z_{i}\omega ^{C_{i}^{a}},
\end{equation}
with $\omega ^{5}=1$ and $C_{i}^{a}$ , to which we refer in this paper as
Calabi-Yau charges, are as follows
\begin{eqnarray}
C_{i}^{1} &=&(1,-1,0,0,0)  \nn \\ C_{i}^{2} &=&(1,0,-1,0,0) \\
C_{i}^{3} &=&(1,0,0,-1,0)\nn.
\end{eqnarray}
These charges satisfy $\sum_{i=1}^{5}C_{i}^{a}=0 $ modulo (5); in
agreement with the CY condition. Note that under the action (3),
the coordinate $z_{5}$ is stable. Using this symmetry, one can
build an orbifold of the quintic by identifying  $z_{i}$  points
related by eqs(3). The next step is to consider a coordinate patch
of $\mathbf{P}^{4}$ where $z_{5}=1$, associate to the $\left\{
z_{1},z_{2},z_{3},z_{4}\right\} $\ local variables, the set of
$5\times 5$\ matrix operators $\left\{
Z_{1},Z_{2},Z_{3},Z_{4}\right\} ;$ $Z_{5}$\ with the $I_{5}$\
identity matrix, and construct the non commutative quintic
associated with eqs(2-4) by solving the commutative quintic
constraint eqs. The result obtained in \cite{15} is given by (1).
As one sees, eqs(1) seems to be a more general result as it does
not depend explicitly on the Calabi-Yau charges $C_{i}^{a}$
eqs(4). In fact eqs(1) define an abstract non commutative algebra
valid for any choice of the set $\left\{ C_{i}^{a}\right\} $\
charges satisfying the Calabi-Yau condition.
\begin{equation}
\sum_{i=1}^{5}C_{i}^{a}=0, \qquad  \mbox {modulo (5) }.
\end{equation}
In what follows we study this property by considering different
choices of sets$\ \left\{ C_{i}^{a}\right\} $\ and show that they
lead indeed to the same non commutative algebra(1). What we want
to show throughout this examination is that different choices of
sets of Calabi-Yau charges lead to different realizations of the
algebra(1) and so constitute different representations of the non
commutative quintic. To that purpose, let us first reconsider the
example (4) and rederive eqs(1) in our manner by solving
constraints on the non commutative quintic structure, then we
study the second set of Calabi-Yau charges $C_{i}^{a}$\

\begin{eqnarray}
C_{i}^{1} &=&(1,-2,1,0,0)   \nn\\ C_{i}^{2} &=&(0,1,-2,1,0) \\
C_{i}^{3} &=&(1,0,1,-2,0)\nn.
\end{eqnarray}
Finally we show how things extend for generic choices of $\left\{
C_{i}^{a}\right\} $ sets\ and give the general result.

\subsection{Representation I}

Using the previous analysis, one can define the local non commutative
quintic around $Z_{5}=z_{5}I$\ as follows
\begin{eqnarray}
Z_{i}Z_{j} &=&\theta _{ij}Z_{j}Z_{i}\qquad i,j=1,...,4,
\\ Z_{i}Z_{5} &=&Z_{5}Z_{i},
\end{eqnarray}
together with the constraint eqs
\begin{eqnarray}
\left[ Z_{j},Z_{i}^{5}\right] &=&0,   \nn\\ \left[
Z_{j},\prod_{i=1}^{4}Z_{i}\right] &=&0.
\end{eqnarray}
The $\theta _{ij}$'s are non zero complex parameters carrying non
commutativity of the algebra ${\mathcal{A}}_{nc}(5)$. As the monomials $%
Z_{j}^{5}$ and $\ \prod_{i=1}^{5}\left( Z_{i}\right) $ commute with all $%
Z_{i}$'s, it follows that the $\theta _{ij}$'s  should obey the
following constraint relations:
\begin{eqnarray}
\theta _{ij}^{5} &=&1, \\ \theta _{ij}\theta _{ji} &=&1,\qquad
\forall i,j \\ \prod_{i=1}^{4}\theta _{ji} &=&1,\qquad \forall j.
\end{eqnarray}
These relations constitute actually the defining conditions of
local non commutative quintic with discrete torsion and show that
the $\theta _{ij}$ parameters carry $\frac{3\times 4}{2}-(4-1)=3$
degrees of freedom in agreement with the discrete torsion eq(2).
Let us comment briefly these constraints eqs(10-12) and show that
the $\theta _{ij}$\ 's are solved as:
\begin{equation}
\theta _{ij}=\exp i\left( \frac{2\pi }{5}L_{ij}\right) =\omega ^{L_{ij}}.
\end{equation}
Eq(11) requires that $L_{ij}$\ is a $4\times 4$\ antisymmetric matrix, i.e $%
L_{ij}=-L_{ji},$ while eq(12) implies $\sum_{i}^{4}L_{ij}=0\;\;
\mbox {modulo (5) }$. The key idea of our way of solving the constraint $%
\sum_{i}^{4}L_{ij}=0 \;\;\mbox {modulo (5) }$ is to interpret it
as just a manifestation of the Calabi-Yau condition (5). This
observation suggests that $L_{ij}$ can be solved by antisymmetric
bilinears of $C_{i}^{a}$ as shown herebelow:
\begin{eqnarray}
L_{ij} &=&m_{12}\left(
C_{i}^{1}C_{j}^{2}-C_{j}^{1}C_{i}^{2}\right) -m_{23}\left(
C_{i}^{2}C_{j}^{3}-C_{j}^{2}C_{i}^{3}\right)  \nn \\
&&+m_{13}\left( C_{i}^{1}C_{j}^{3}-C_{j}^{1}C_{i}^{3}\right),
\end{eqnarray}
where $m_{12}=k_{1}$, $m_{23}=k_{2}$ and $m_{13}=k_{3}$\ are
integers modulo $5$. In other words
\begin{equation} L_{ij}=\left(
\begin{array}{cccc}
0 & k_{1}-k_{3} & -k_{1}+k_{2} & k_{3}-k_{2} \\
-k_{1}+k_{3} & 0 & k_{1} & -k_{3} \\
k_{1}-k_{2} & -k_{1} & 0 & k_{2} \\
-k_{3}+k_{2} & k_{3} & -k_{2} & 0
\end{array}
\right) .
\end{equation}
Therefore the non commutative quintic associated to eqs(2-4),
reads as:
\begin{eqnarray}
Z_{1}Z_{2} &=&\omega ^{k_{1}-k_{3}}Z_{2}Z_{1},\qquad
Z_{1}Z_{3}=\omega ^{-k_{1}+k_{2}}Z_{3}Z_{1},   \nn\\ Z_{1}Z_{4}
&=&\omega ^{k_{3}-k_{2}}Z_{4}Z_{1},\qquad Z_{2}Z_{3}=\omega
^{k_{1}}Z_{3}Z_{2}, \\ Z_{2}Z_{4} &=&\omega
^{-k_{3}}Z_{4}Z_{2},\qquad Z_{3}Z_{4}=\omega
^{k_{2}}Z_{4}Z_{3}\nn.
\end{eqnarray}
Setting $\alpha =\omega ^{k_{1}-k_{3}},$ $\beta =\omega ^{k_{1}-k_{2}}$ and $%
\gamma =\omega ^{k_{3}}$, one discovers that these relations are
identical to the relations (1); so eqs(16) define indeed a
representation of the non commutative algebra
${\mathcal{A}}_{nc}(5)$.\\ Eqs(16) may realized in terms of
$5\times 5$\ matrices in various ways; one of them which was
studied in \cite{15} is based on fixing $\beta $\ and $\gamma $\
in terms of $\alpha $\ ($\beta =\alpha ^{n+1},\gamma =\alpha
^{m+1}$) breaking by the way $\mathbf{Z}_{5}^{3}$\ down to
$\mathbf{Z}_{5}.$ Here we give a solution preserving manifestly
the $\mathbf{Z}_{5}^{3}$\ symmetry. This solution involves the
following $5\times 5$\ matrices
\begin{equation}
{\bf P}_{\eta }=diag(1\mathbf{,\eta ,}\eta ^{2},\eta ^{3},\eta
^{4}),\quad {\bf Q}=\left(
\begin{array}{ccccc}
0 & 0 & 0 & 0 & 1 \\
1 & 0 & 0 & 0 & 0 \\
0 & 1 & 0 & 0 & 0 \\
0 & 0 & 1 & 0 & 0 \\
0 & 0 & 0 & 1 & 0
\end{array}
\right)
\end{equation}
where $\eta \ $stands for $\omega ^{k_{1}},\omega ^{k_{2}}$,\
$\omega ^{k_{3}}$ and their products.  In terms of these matrices,
the $Z_{i}$ 's\ read, up to a normalization factor, as:
\begin{eqnarray}
Z_{1} &=&z_{1}{\bf P}_{\omega ^{k_{1}+k_{2}+k_{3}}}{\bf Q}^{3}
\nn\\ Z_{2} &=&z_{2}{\bf P}_{\varpi ^{k_{1}}}{\bf Q}^{-1}  \nn \\
Z_{3} &=&z_{3}{\bf P}_{\varpi ^{k_{2}}}{\bf Q}^{-1} \\ Z_{4}
&=&z_{4}{\bf P}_{\varpi ^{k_{3}}}{\bf Q}^{-1}\nn,
\end{eqnarray}
where $\varpi $ is the complex conjugate of $\omega$. This
solution, which
extends that given in \cite{15}, shows clearly that $Z_{i}^{5}$\ and the product $%
\prod_{i=1}^{4}Z_{i}$\ are in the centre $\mathcal{Z}$ of the
local non commutative algebra ${\mathcal{A}}_{nc}(5).$ Actually
these relations define the ${\mathcal{R}}_{\left\{
C_{i}^{a}\right\} }$ representation of eqs(1). \\ We end this
paragraph by making three remarks: $(i)$ By an appropriate choice
of the values of the $k_{1},k_{2}$\ and $k_{3}$\ integers on can
get the subalgebras of ${\mathcal{A}}_{nc}(5).$ For example,
taking $k_{3}=0,$ one gets the following subalgebra
\begin{eqnarray}
Z_{1}Z_{2} &=\omega ^{k_{1}}Z_{2}Z_{1},\qquad Z_{1}Z_{3}&=\omega
^{-k_{1}+k_{2}}Z_{3}Z_{1},   \nn\\ Z_{1}Z_{4} &=\omega
^{-k_{2}}Z_{4}Z_{1},\qquad Z_{2}Z_{3}&=\omega ^{k_{1}}Z_{3}Z_{2},
\\ Z_{2}Z_{4} &=Z_{4}Z_{2},\qquad Z_{3}Z_{4}&=Z_{4}Z_{3}\nn.
\end{eqnarray}
Moreover if we take $k_{2}=k_{3}=0,$\ the algebra (16) reduces to
\begin{eqnarray}
Z_{1}Z_{2} &=&\omega ^{k_{1}}Z_{2}Z_{1},\qquad Z_{1}Z_{3}=\omega
^{-k_{1}}Z_{3}Z_{1},\qquad Z_{2}Z_{3}=\omega ^{k_{1}}%
Z_{3}Z_{2},   \\ Z_{1}Z_{4} &=&Z_{4}Z_{1},\qquad
Z_{2}Z_{4}=Z_{4}Z_{2},\qquad Z_{3}Z_{4}=Z_{4}Z_{3}\nn.
\end{eqnarray}
This subalgebra could be linked to the non commutative K3 algebra
studied in \cite{16}. $(ii)$ The solution we have given in eq(14)
is a particular one. A more general solution involving the toric
geometry data of Calabi-Yau manifolds as well as other features
can be found in \cite{17}. $(iii)$ Viewing the algebra eq(1) as
describing a  ${\bf {Z}}_{5}^{2}$\ orbifold of the quintic with a
${\bf Z}_{5}$ discrete torsion and considering the singularities
in codimension two, it was shown in [15] that branes do
fractionate at the orbifold points. In subsection 4.2, we will
give explicit details for the example of the ${\bf Z}_{8}^{6}$
orbifold of the {\it eight-tic}.

\subsection{Representation II}

Here we give an other representation of the algebra (1) by using
Calabi-Yau charges chosen as in eqs(6). As one sees, these charges
are different from those given by eq(4); they are rather similar
to the toric data of the blown up of the ${\hat A}_2$ affine
singularity of the ALE space. Eqs(6) are used in geometric
engineering of \ fundamental matters in $4D\ N=2 $ superconformal
theories \cite{18,19}. An analogous analysis as that used in the
previous subsection leads to:
\begin{equation}
L_{ij}=\left(
\begin{array}{cccc}
0 & L_{12} & L_{13} & L_{14} \\
-L_{12} & 0 & L_{23} & L_{24} \\
-L_{13} & -L_{23} & 0 & L_{34} \\
-L_{14} & -L_{24} & -L_{34} & 0
\end{array}
\right)
\end{equation}
with
\begin{eqnarray}
L_{12} &=&k_{1}+k_{2}+2k_{3},\quad L_{13}=-2k_{1}-2k_{2}, \nn  \\
L_{14} &=&k_{1}+k_{2}-2k_{3},\quad L_{23}=3k_{1}-k_{2}-2k_{3}, \\
L_{24} &=&-2k_{1}+2k_{2}+4k_{3},\quad
L_{34}=k_{1}-3k_{2}-2k_{3},\nn
\end{eqnarray}
and where $k_{1},k_{2}$\ and $k_{3}$\ are integers modulo $5.$ The
new local algebra describing the non commutative quintic reads now
as:
\begin{eqnarray}
Z_{1}Z_{2} &=Z_{2}Z_{1}\omega ^{k_{1}+k_{2}+2k_{3}} ,\quad
Z_{1}Z_{3}&=Z_{3}Z_{1}\omega ^{-2k_{1}-2k_{2}},\quad  \nn\\
Z_{1}Z_{4} &=Z_{4}Z_{1}\omega ^{k_{1}+k_{2}-2k_{3}},\quad
Z_{2}Z_{3}&=Z_{3}Z_{2}\omega ^{3k_{1}-k_{2}-2k_{3}},\quad  \\
Z_{2}Z_{4} &=Z_{4}Z_{2}\omega ^{-2k_{1}+2k_{2}+4k_{3}},\quad
Z_{3}Z_{4}&=Z_{4}Z_{3}\omega ^{k_{1}-3k_{2}-2k_{3}}.\nn
\end{eqnarray}
Setting $\alpha =\omega ^{k_{1}+k_{2}+2k_{3}},$ $\beta =\omega
^{2k_{3}-k_{1}-k_{2}}$ and $\gamma =\omega
^{2k_{1}-2k_{2}-4k_{3}}$, one discovers once again the non
commutative algebra (1).\ Here also one can derive the various
subalgebras of eqs(23) by appropriate choices of the \ integers
$k_{1},k_{2}$\ and $k_{3}$. The corresponding subalgebras define
special non commutative geometries. To conclude this section, one
should retain that\ eqs (16) and (23) define two realizations of
the local noncommutative quintic  algebra. As such, different
choices of sets $\{C_{i}^{a}\}$ of Calabi-Yau charges lead to
different realizations of the local algebra
${\mathcal{A}}_{nc}(5).$\ Moreover the special solutions of eq(14)
give the subalgebras of eq(1). Such analysis can be extended to
other Calabi-Yau manifolds with discrete torsion. For more details
see \cite{17}.

\subsection{Other Representations}

In this paragraph we want to give a comment on other possible
realisations of the non commutative algebra (1) using orbifolds of
hypersurfaces in the weighted projective spaces $\mathbf{WP}^{4}.$
To that purpose recall that in the weighted projective space
$\mathbf{WP}_{\{\delta _{1},\delta _{2},\delta _{3},\delta
_{4},\delta _{5}\}}^{4},$ the analogue of eq(2) reads as
\begin{equation}
\sum_{\i =1}^{5}u_{i }^{\frac{D}{\delta _{i }}}+a_{0}\prod_{i
=1}^{5}\left( u_{i }\right) =0.
\end{equation}
where $D=\sum_{i}\delta _{i}.$ To fix the ideas, let us consider
the simple example where $\delta _{1}=2$ while $,\delta
_{2}=\delta _{3}=\delta _{4}=\delta _{5}=1.$ In this case, the
analogue of the quintic polynomial becomes:
\begin{equation}
u_{1}^{3}+u_{2}^{6}+u_{3}^{6}+u_{4}^{6}+u_{5}^{6}+a_{0}\prod_{i=1}^{5}\left(
u_{i}\right) =0.
\end{equation}
where $(u_{1},u_{2},u_{3},u_{4},u_{5})$ are the quasi-homogeneous
coordinates of $WP_{\{2,1,1,1,1\}}^{4}$ and $a_{0}$ is a complex parameter.
This polynomial has also a set of discrete isometries acting on the
homogeneous coordinates $u_{i}$ as:
\begin{equation}
u_{i}\rightarrow u_{i}\zeta _{i}^{C_{i}^{a}}\qquad i=1,...,5
\end{equation}
with $\zeta _{1}^{3}=1$ while the others $\zeta _{i}^{6}=\omega ^{6}=1;$ i.e
$\zeta _{1}=\omega ^{2}$ and $\zeta _{i}=\omega ,\ i=2,..,5$.  The $C_{i}^{a}
$ are chosen as in eqs(4). Note that $u_{5}$ is stable under the change(26).
Using this symmetry and following the non commutative algebraic method, one
can build an orbifold of this Calabi-Yau threefolds (25) by identifying $%
u_{i}$\ points\ rotated by eqs(26). In the coordinate patch where
$u_{5}=1$, the four variables $\left\{
u_{1},u_{2},u_{3},u_{4}\right\} $ obey, after performing the
correspondence, the non commutative relations.

\begin{equation}
U_{i}U_{j}=\theta _{ij}\ U_{j}U_{i},
\end{equation}
where the $\theta _{ij}$'s are non zero complex parameters.\ As
the monomials $U_{1}^{3},$ $U_{i}^{6}$ and
$\prod\limits_{i=1}^{5}\left( U_{i}\right) $ are in the centre
$\mathcal{Z}$ of the algebra, we have:
\begin{eqnarray}
\left[ U_{j},U_{1}^{3}\right]  &=&0, \nn \\ \left[
U_{j},U_{i}^{6}\right]  &=&0\qquad i=2,3,4, \\ \left[
U_{j},\prod_{i=1}^{4}U_{i}\right]  &=&0\nn.
\end{eqnarray}
Therefore the $\theta _{ij}$'s  should obey the following
constraints:

\begin{eqnarray}
\theta _{i1}^{3} &=&1,\qquad =2,3,4, \\ \theta _{ij}^{6}
&=&1,\qquad j\neq 1,i \\ \prod_{i=1}^{4}\theta _{ji} &=&1,\qquad
\forall j \\ \theta _{ij}\theta _{ji} &=&1,\qquad \forall i,j.
\end{eqnarray}
The solution of these constraints are built as follows: first use eqs
(29-30) and (32) to write $\theta _{ij}$ as
\begin{eqnarray}
\theta _{ij} &=&\exp i\left( \frac{2\pi }{6}L_{ij}\right) =\omega
^{L_{ij}},\qquad L_{ij}=-L_{ji},\qquad \nn  \\ L_{i1}
&=&0,2,4\qquad \mbox{modulo 6}.
\end{eqnarray}
Eqs (31) implies however that $\sum_{i}^{4}L_{ij}=0\;\;\mbox
{modulo (6) }$. Particular solutions may be learned from eqs(14)
using antisymmetric
bilinears of $C_{i}^{a}$ (4). Straightforward calculations show that $L_{ij}$%
\ is given by the following $4\times 4$ matrix:
\begin{equation}
L_{ij}=\left(
\begin{array}{cccc}
0 & k_{1}-k_{3} & -k_{1}+k_{2} & k_{3}-k_{2} \\
-k_{1}+k_{3} & 0 & k_{1} & -k_{3} \\
k_{1}-k_{2} & -k_{1} & 0 & k_{2} \\
-k_{3}+k_{2} & k_{3} & -k_{2} & 0
\end{array}
\right)
\end{equation}
where now  the $k_{i}$\ integers are such that
\begin{equation}
k_{i}-k_{j}\equiv 2r_{ij}\in 2Z.
\end{equation}
 Therefore the non commutative algebra associated to eq(25) is
\begin{eqnarray}
U_{1}U_{2} &=&\omega ^{k_{1}-k_{3}}U_{2}U_{1},\quad
U_{1}U_{3}=\omega ^{k_{2}-k_{1}}U_{3}U_{1},  \nn \\ U_{1}U_{4}
&=&\omega ^{k_{3}-k_{1}}U_{4}U_{1},\qquad U_{2}U_{3}=\omega
^{k_{1}}U_{3}U_{2}, \\
U_{2}U_{4} &=&\omega ^{-k_{3}}U_{4}U_{2},\qquad %
U_{3}U_{4}=\omega ^{k_{2}}U_{4}U_{3}.\nn
\end{eqnarray}
Moreover taking $\ \alpha =\omega ^{k_{1}-k_{3}},$ $\beta =\omega
^{k_{1}-k_{3}}$ and $\gamma =\omega ^{k_{3}}$, one discovers the non
commutative algebra (1),\ except now the deformation parameters are as
follows:
\begin{equation}
\alpha ^{3}=\beta ^{3}=\gamma ^{6}=1.
\end{equation}
Note that for the hypersurface (24) in the weighted projective
space $\mathbf{WP} _{\{1,2,1,1,1\}}^{4}$ namely
\begin{equation}
u_{1}^{6}+u_{2}^{3}+u_{3}^{6}+u_{4}^{6}+u_{5}^{6}+a_{0}\prod_{i
=1}^{5}\left( u_{i}\right) =0,
\end{equation}
with the discrete symmetries $u_{i}\rightarrow u_{i}$ $\zeta
_{i}^{C_{i}^{a}};$ where $\zeta _{2}=\omega ^{2}$ and $\zeta
_{i\neq 2}=\omega \ $and where the $C_{i}^{a}$ are as in eqs(4),
the analogue of the constraint eqs (36) are given by $k_{1}\in 2Z$
and $k_{3}\in 2Z$. Such analysis generalizes naturally  other
situations.

\section{Non Commutative Calabi-Yau}

\qquad The analysis we have developed for the non commutative quintic (2)
and the hypersurfaces (24-25) can be extended straightforwardly to more
general $d$ complex dimension Calabi-Yau homogeneous hypersurfaces in $%
\mathbf{P}^{d+1}$ and $\mathbf{WP}^{d+1}$.  Starting from $(d+2)-tic$ in $\mathbf{P}^{d+1} \cite{20,21};$%
\begin{equation}
z_{1}^{d+2}+z_{2}^{d+2}+z_{3}^{d+2}+z_{4}^{d+2}+z_{5}^{d+2}+\ldots+z_{d+2}^{d+2}+a_{0}\prod_{i
=1}^{d+2}z_{i }=0,
\end{equation}
with the discrete isometries type (3) with Calabi-Yau charges\
$C_{i}^{a}$ charges satisfying the condition $$
\sum_{i=1}^{d+2}C_{i}^{a}=0,\qquad a=1,...,d,$$
one can build a $(d+1)\times (d+1)$ antisymmetric matrix in terms of $%
C_{i}^{a}$\ bilinears as:
\begin{equation}
L_{ij}=-L_{ji}=m_{ab}C_{i}^{[a}C_{j}^{b]},
\end{equation}
where $m_{ab}$\ is an antisymmetric\ $d\times d$ matrix of integers modulo ($%
d+2),$ satisfying
\begin{equation}
\sum_{i=1}^{d+2}L_{ij}=0.
\end{equation}
The non commutative extension of eq(39) is given by the following algebra,
to which we refer to as ${\mathcal{A}}_{nc}(d+2);$%
\begin{eqnarray}
Z_{i}Z_{j} &=&\omega _{ij}\varpi _{ji}Z_{j}Z_{i};\qquad \
i,j=1,...,(d+1),   \nn\\ Z_{i}Z_{d+2} &=&Z_{d+2}Z_{i};\qquad \
i=1,...,(d+1),
\end{eqnarray}
where $\varpi _{kl}$ is the complex conjugate of $\omega _{kl}.\ $\ The non
commutativity parameters are realized in terms of the Calabi-Yau charges as
follows:
\begin{equation}
\omega _{ij}=\exp i\left( \frac{2\pi }{d+2}m_{ab}C_{i}^{a}C_{j}^{b}\right)
=\omega ^{m_{ab}C_{i}^{a}C_{j}^{b}}.
\end{equation}
 Note that this representation involve $d(d-1)/2$ integers
$m_{[ab]}$ in agreement with the numbers of degrees of freedom one
gets using the generalization of constraint eqs(10-12). Note also
that the huge form of the non commutative algebra(42) describing
the non commutative version of eq(39) has $d(d-1)/2$ free
parameters generating the ${\mathbf{Z}}_{d+2}^{d(d-1)/2}$\
discrete symmetry group. By appropriate choices of the \ integers,
one gets all possible subalgebras describing non commutative
special geometries associated with eq(39). For example there are
$d(d-1)/2$ types of\ non commutative subalgebras with discrete
symmetry ${\mathbf{Z}}
_{d+2}^{(d-1)(d-2)/2}$\ and $\frac{d(d-1)}{2}\left( \frac{d(d-1)}{2}%
-1\right) $ subsubalgebras with
${\mathbf{Z}}_{d+2}^{(d-2)(d-3)/2}$ symmetry and so on. Note
finally that for $d=2$ describing K3, eqs(42) reduce to
\begin{eqnarray}
Z_{1}Z_{2} &=&\alpha Z_{2}Z_{1};\quad Z_{1}Z_{3}=\overline{\alpha }%
Z_{3}Z_{1}   \nn\\
Z_{2}Z_{3} &=&\alpha Z_{3}Z_{2};\quad Z_{i}Z_{4}=%
Z_{4}Z_{i};\qquad \ i=1,2,3.
\end{eqnarray}
where we have set $\alpha =\omega _{12}$ $\varpi _{12}=\omega _{23}$ $\varpi
_{23}=$ $\omega _{31}$ $\varpi _{31}$ and $\overline{\alpha }$ is its
complex conjugate.

\section{Matrix Representation of the $Z_{i}$'s}

Finite dimensional representations of the non commutative algebra
are given\ by matrix  subalgebras ${\mathcal{M}}at\left[
n(d+2),C\right] $, the algebra of $n(d+2)\times n(d+2)$ complex
matrices, with $n=1,2,...$. To see this property it is enough to
take the determinant of eqs(42), which constraint the dimension
$D$ of the representation to be such that:
\begin{equation}
\left( \omega _{ij}\varpi _{ji}\right) ^{D}=1.
\end{equation}
Using the identity (43), one discovers that $D$ is a multiple of
$(d+2)$. In what follows we consider the fundamental $(d+2)\times
(d+2)$ matrix representation obtained by extending the solution
(17-18). We will distinguish two situations according to whether
we are dealing with regular or singular points of the Calabi-Yau
orbifold. \subsection{Representations for regular points}
Introducing the following set $\left\{ {\bf{Q;P}}_{\alpha
_{ab}};a,b=1,...,d\right\} $\ of matrices:
\begin{equation}
{\bf P}_{\alpha _{ab}}={diag(1,\alpha _{ab},}\alpha
_{ab}^{2},...,\alpha _{ab}^{d+1});\quad {\bf Q}=\left(
\begin{array}{ccccccc}
0 & 0 & 0 & . & . & . & 1 \\
1 & 0 & 0 & . & . & . & 0 \\
0 & 1 & 0 & . & . & . & 0 \\
. & . & . & . & . & . & . \\
. & . & . & . & . & . & . \\
0 & 0 & 0 & . & 1 & 0 & 0 \\
0 & 0 & 0 & . & . & 1 & 0
\end{array}
\right)
\end{equation}
satisfying ${\bf P}_{\alpha }^{d+2}=1,$ ${\bf P}_{\alpha }{\bf P}%
_{\beta }={\bf P}_{\alpha \beta }$\ \ and $ {\bf Q}^{d+2}=1;$ then
setting
\begin{eqnarray}
Z_{1} &=&z_{1}\prod_{a,b=1}^{d}\left( {\bf P}_{\alpha _{ab}}^{C_{1}^{a}}%
 {\bf Q}^{C_{1}^{b}}\right)   \nn \\
Z_{2} &=&z_{2}\prod_{a,b=1}^{d}\left( {\bf P}_{\alpha _{ab}}^{C_{2}^{a}}%
 {\bf Q}^{C_{2}^{b}}\right) ,
\end{eqnarray}
and computing $Z_{1}Z_{2}$\ and $Z_{2}Z_{1,}$\ one discovers $$
Z_{1}Z_{2}=\eta _{12}Z_{2}Z_{1}, $$ with $\eta _{12}$ equal to
\begin{equation}
\eta _{12}=\omega ^{m_{[ab]}C_{1}^{a}C_{2}^{b}}.
\end{equation}
Note that this relation is a nothing else the leading term of a
more general tensor given by:
\begin{equation}
\eta _{ij}=\prod_{a,b}\left( \left[ \alpha _{ab}\overline{\alpha }%
_{ba}\right] ^{C_{i}^{a}C_{j}^{b}}\right) ,
\end{equation}
where $\alpha _{ab}=\omega ^{m_{ab}}$. The remaining others can be
derived as powers of the $ {\bf P}_{\alpha _{ab}}$and $ {\bf Q}$
matrices so that the product $\prod_{i=1}^{d+1}Z_{i}$\ is in the
centre $\mathcal{Z}$ of the local non commutative algebra
${\mathcal{A}}_{nc}\left[ d+2\right]$.  Thus the solution  for the
$Z_{i}^{\prime }s$ \ reads, up to a normalization factor, as
\begin{equation}
Z_{i}=z_{i}\prod_{a,b=1}^{d}\left( {\bf P}_{\alpha
_{ab}}^{C_{i}^{a}} {\bf Q}^{C_{i}^{b}}\right).
\end{equation}
One can check easily  that eqs(50)  satisfy naturally
$Z_{i}^{d+2}\sim {\bf I}_{d+2}$,  the Calabi-Yau condition
\begin{eqnarray}
\prod_{i=1}^{d+1}Z_{i} &=&\prod_{a,b=1}^{d}\left[ \prod_{i=1}^{d+1}\left(
z_{i}\mathbf{P}_{\alpha _{ab}}^{C_{i}^{a}} {\bf Q}%
^{C_{i}^{b}}\right) \right]  \nn \\ &=&{\bf I}_{d+2}\left(
\prod_{i=1}^{d+1}z_{i}\right) .
\end{eqnarray}
and so are indeed solutions of the non commutative algebra (42),
with
\begin{equation}
\eta _{ij}=\omega _{ij}\varpi _{ji}.
\end{equation}
More details on this derivation as well as other features
involving toric geometry of Calabi-Yau manifolds are given in
\cite{17}. \subsection{Fractional Branes} The solutions we gave
earlier correspond to regular points of non commutative
Calabi-Yau. The ${\cal{A}}_{nc}(d+2)$ representations eq(51) are
irreducibles and the branes do not fractionate. Similar solutions
may be worked out as well for orbifold points with discrete
torsions where we expect to get fractional branes. Due to the
richness of possibilities, we will focus our attention herebelow
on giving a particular solution for the ${\bf {Z}}_{8}^{6}$
orbifold of the {\it{eight-tic}}; general solutions will be
reported in [17], see also [22] for  the resolution of stringy
singularities by non commutative algebras. To that purpose
consider the orbifold of eq(39), $d=6,$ with respect to ${\bf
{Z}}_{8}^{6}$ and take the Calabi-Yau vector charges as:
\begin{eqnarray}
C_{i}^{1} &=&(1,-1,0,0,0,0,0,0)  \nn \\ C_{i}^{2}
&=&(1,0,-1,0,0,0,0,0) \nn\\ C_{i}^{3} &=&(1,0,0,-1,0,0,0,0)\nn\\
C_{i}^{3} &=&(1,0,0,0,-1,0,0,0)\\
 C_{i}^{3} &=&(1,0,0,0,0,-1,0,0)\nn\\
 C_{i}^{3} &=&(1,0,0,0,0,0,-1,0).\nn
\end{eqnarray}
For regular points, the matrix representation of eqs(42) is
irreducible as shown on eqs (51). For fixed points however, the
situation is more subtle as there exists situations where
representations are reducible. One way to deal with the
singularity of the orbifold with respect to ${\bf {Z}}_{8}^{6}$ is
to interpret the algebra as describing a ${\bf Z}_{8}^{3}$
orbifold with ${\bf Z}_{8}^{3}$\ discrete torsions having
singularities in codimension four. Starting from eqs(42) and
choosing $Z_{5}$, $Z_{6}$ and $Z_{7}$ in the centre of the algebra
by setting
\begin{equation}
\left( \omega _{ij}\varpi _{ji}\right) =1,\quad  \mbox{
for}\;\;i=5,6,7,8;\qquad \forall j=1,...,8,
\end{equation}
the algebra reduces to \bea
 Z_{1}Z_{2} &=&\alpha _{1}\alpha
_{2}Z_{2}Z_{1}\nn\\ Z_{1}Z_{3}&=&\alpha _{1}^{-1}\alpha
_{3}Z_{3}Z_{1}\nn\\ Z_{1}Z_{4} &=&\alpha _{2}^{-1}\alpha
_{3}^{-1}Z_{4}Z_{1}\\
 Z_{2}Z_{3}&=&\alpha _{1}Z_{3}Z_{2}\nn\\
 Z_{2}Z_{4}
&=&\alpha _{2}Z_{4}Z_{2},\nn\\ Z_{3}Z_{4}&=&\alpha
_{3}Z_{4}Z_{3}\nn \eea
 and all remaining other relations are commuting. In this
equation, the $\alpha _{i}$ 's are such that $\alpha _{i}^{8}=1;$
these are the phases of the ${\bf Z}_{8}^{3}$ discrete torsions. \
In the singularity where the $z_{1}$, $z_{2},$\ $z_{3},$ and
$z_{4}$ moduli of eq(51) go to zero, one ends with the familiar
result for orbifolds with discrete torsion. Therefore the branes
fractionate in the codimension four singularities. Actually this
result extends naturally the one on the non commutative quintic
obtained in [15] and can be generalized straightforwardly [17].
\section{Conclusion}

In this paper we have studied the building of non commutative
Calabi-Yau hypersurface orbifolds using the algebraic geometry
approach of \cite{15} and the toric data of the Calabi-Yau
manifolds with discrete torsion. Our main results are summarized
as follows: $\mathit{(i)}$ We have reviewed the construction of
the non commutative quintic ${\mathcal{A}}_{nc}(5)$, its
subalgebras and built new representations using methods of toric
geometry. We have also studied the analogue of the non commutative
quintic in the weighted projective space $\mathbf{WP}^{4}$.
$\mathit{(ii)}$ We have given the the generalization of these
results to higher dimensional Calabi-Yau hypersurface orbifolds
and derived the explicit form of the corresponding non commutative
algebra ${\mathcal{A}}_{nc}(d+2)$ and its bulk and orbifold
representations. As an example we  have studied the fractional
branes for the ${\bf Z}_{8}^{3}$ orbifold of the {\it eight-tic}
with ${\bf Z}_{8}^{3}$\ discrete torsions. One of the lessons we
learnt in this work is that to each Calabi-Yau data $ \left\{
C_{i}^{a}\right\} $ corresponds a  representation ${\mathcal{R}}
_{\left\{ C_{i}^{a}\right\} }$ of the non commutative algebra
${\mathcal{A}}_{nc}(d+2)$.
\\ \\ {\bf
Acknowledgments}\\ Adil Belhaj would like to thank the organizers
of the Spring Workshop on Superstrings and related Matters (2001),
the Abdus Salam International Centre for  Theoretical Physics,
Trieste, Italy, for hospitality.  He would also  like   to thank
D. Berenstein,  R. Gopakumar and S. Theisen  for  discussions.
This work is supported by SARS, programme de soutien \`a la
recherche scientifique; Universit\'e Mohammed V-Agdal, Rabat.\\

\end{document}